**Title:** Monatomic phase change memory

**Authors:** Martin Salinga*[1,2], Benedikt Kersting[1,2], Ider Ronneberger[1,2], Vara Prasad Jonnalagadda[1], Xuan Thang Vu[2], Manuel Le Gallo[1], Iason Giannopoulos[1], Oana Cojocaru-Mirédin[2], Riccardo Mazzarello[2], Abu Sebastian[1]

[1] IBM Research-Zurich, CH-8803 Rüschlikon, Switzerland

[2] RWTH Aachen University, D-52074 Aachen, Germany

**Phase Change Memory (PCM) has been developed into a mature technology capable of storing information in a fast and non-volatile way[1-3], with potential for neuromorphic computing applications[4-6]. However, its future impact in electronics depends crucially on how the materials at the core of this technology adapt to the requirements arising from continued scaling towards higher device densities. A common strategy to fine-tune the properties of PCM materials, reaching reasonable thermal stability in optical data storage, relies on mixing precise amounts of different dopants, resulting often in quaternary or even more complicated compounds[6-8]. Here we show how the simplest material imaginable, a single element (in this case antimony), can become a valid alternative when confined in extremely small volumes. This compositional simplification eliminates problems related to unwanted deviations from the optimized stoichiometry in the switching volume, which become increasingly pressing when devices are aggressively miniaturized[9,10]. Removing compositional optimization issues may allow capitalizing on nanosize effects in information storage.**

Continued miniaturization of PCM devices is not only required to achieve memory chips with higher data density and neuromorphic hardware capable of processing larger amounts of information, but also to increase the power efficiency per operation. PCM is based on the ability to switch so-called phase change materials between crystalline states with high electrical conductivity and meta-stable amorphous states with low electrical conductivity. The electrical pulses thermally inducing the fast transitions between those states can be of less energy when less material needs to be heated up[11,12]. In the endeavour to scale further and further one is reaching volumes in which it starts to make a significant difference to the composition in the switchable regime, whether there are a few more atoms of the right kind in the ensemble or not. And even if we manage to achieve the precise stoichiometry in each of the billions of PCM elements across a silicon wafer, it is known that under the regularly applied strong electric fields (> 0.1 V/nm) and high temperatures (> 1000 K) the different elements move in and out of the region of interest – by field-assisted motion or simply by phase segregation – limiting the device cyclability and lifetime[9,10]. This also results in significant increase in stochasticity associated with the operation of these devices. Another key challenge with miniaturization is that the interface effects become increasingly important. Various studies have shown how the crystallization kinetics of phase change materials, when scaled into few nanometer length scales, can change significantly depending on the material with which it is in contact[13-16]. Thus, compositions carefully trimmed for showing ideal characteristics in larger volumes must be expected to lose their favourable behaviour in aggressively scaled devices.

In this work, we take a drastically different approach. Instead of tweaking the properties of phase change materials by compositional variation, we decide to use the naturally most homogeneous and at the same time simplest material imaginable: a pure element. Amorphous Sb has been found to conduct electrical charge in a semiconducting way with orders of magnitude higher electrical resistivity than crystalline antimony[17] — a contrast even sufficient for assigning multiple resistance levels per memory cell[18]. However, to-date amorphous Sb could only be created by careful deposition of very thin films or at reduced temperatures[17,19]. The creation of a glass by quenching from the melt, the essential process in switching a PCM, has never been accomplished for pure Sb due to its extreme proneness to crystallization. Recently, however, great progress has been made in forming glasses even out of materials that for the longest time had been deemed impossible to stabilize in a glassy state: most notably, small volumes of monatomic metallic glasses were formed by attaining ultrafast cooling from the melt[20-22].

Building on those findings, in our present study, we first used ab-initio molecular dynamics (AIMD) simulations to study the effect of quenching molten antimony down to room temperature at different cooling rates (more information in methods section). The results show that the stability of Sb against crystallization at room temperature is significantly dependent on the rate at which it was cooled from the melt (Fig. 1). The demonstrated qualitative trend can be expected to hold for much longer times even for more accessible quenching rates. (The transferability of those results to timescales of more practical interest is discussed in the methods section.) Therefore, our AIMD simulations can be taken as encouragement in the sense that if we manage to quench antimony rapidly enough, we can succeed in creating a glass of pure antimony with sufficient stability against crystallization even around room temperature.

Accordingly, we performed experiments using electrically contacted structures of nanometric dimensions (depicted in Fig. 2a) in order to be able to effectively dissipate the heat from the small volume of molten Sb through its interfaces into the surrounding matter (the device fabrication is described in the methods section). Trapezoidal voltage pulses were applied to the devices at various base temperatures (see section on electrical testing at cryogenic temperatures in the methods section for details). Control of the trailing edges of such pulses in the range of few nanoseconds proved to be instrumental (see supplementary information section 'Critical cooling rate for glass formation'). This way, we were able to reliably melt-quench the antimony into a semiconducting state reflected by an increase of the device resistance of more than 2.5 orders of magnitude compared to the fully crystalline state (Fig. 2b), in line with what has been reported in literature for thin films of amorphous Sb deposited on a cold substrate[17]. This contrast in resistance is not only crucial for applications in the field of electronic information processing such as those mentioned above, in a fundamental study like ours it allows for immediate insights into how successful an amorphization attempt has been. The melt-quenched amorphous state can, for example, be observed to exhibit a characteristic temperature dependence of its electrical transport (Fig. 2b). Another attribute of amorphous phase change materials is a temporal increase in electrical resistivity ascribed to structural relaxation of the material towards an energetically more favourable ideal glass state[23,24]. At constant ambient temperature the resistance typically exhibits a temporal dependence characterized by $R(t) = R(t_0) \cdot \left( \frac{t}{t_0} \right)^{\nu}$ where $R(t_0)$ is the resistance measured at time $t_0$. We measured drift coefficients $\nu$ for the melt-quenched amorphous Sb (Fig. 2c) in the range of $0.10 \pm 0.02$, which is remarkably similar to that reported for conventional, multi-elemental phase change materials such as $Ge_2Sb_2Te_5$ [25]. Also, we observe the typical threshold switching phenomenon when measuring the dynamic current response to a voltage sweep (see supplementary information section 'Observation of threshold switching').

Following this examination of evidence of the realization of amorphous Sb, we now turn towards the influence of quench rates through a systematic modification of the trailing edges of the applied electrical pulses (Fig. 2d). At a given peak-power of a trapezoidal voltage pulse applied to a fully crystalline device under test, the resulting high device resistance decreases quite linearly with increasing trailing edge duration of the pulse. While the peak-power determines how much antimony is molten just before the voltage is decreased, the trailing edge controls how far the crystal structure can grow back into the previously molten volume. The more time is spent at elevated temperatures where crystal growth is extremely fast, the smaller is the volume of Sb that survives in a disordered state when ambient temperature is reached again. A trailing edge of only 10 nanoseconds is enough to form a fully crystalline conduction path through the device even in the most ideal of all investigated conditions (lowest ambient temperature). With lower programming powers and thus smaller molten volumes to start the quenching process from, the trailing edges must be even sharper to leave an amorphous plug that can significantly obstruct the charge transport through the device. Further reduction of the programming power must ultimately result in a situation where the molten material does not even cover the whole cross-section of the device, and hence a glass that completely blocks the conduction path cannot be created. These scenarios define the borders of the amorphization window, i.e. the range of programming powers and trailing edges allowing a successful creation of an amorphous mark that is detectable as a pronounced elevation of the device resistance above its crystalline level (Fig. 3 and supplementary information section 'Definition of an amorphization window').

Note that, even if the heating from the electrical pulse could be turned off infinitely quickly, the dissipation of heat from the molten antimony into its surrounding keeps quenching rates finite. In our test structures the thermal boundary conditions of the antimony are strongly dependent on the thickness of the underlying $SiO_2$ acting as a heat barrier towards the vast thermal bath of the silicon wafer. With a 200 nm thin sheet of $SiO_2$ electrically insulating the antimony from the silicon substrate it is barely possible to amorphize any fraction of the Sb, even at an ambient temperature of only 100 K (Fig. 3a). Reducing the thickness of this dielectric layer from 200 nm to 100 nm and further to 40 nm results in substantially faster cooling rates. At the same time, the maximum temperatures reached across the antimony line are reduced by the more effective heat flux into the substrate and so is the size of the molten volume. This reduction, however, is clearly more than counterbalanced by a crucial shortening of the 'thermal trailing edge', which gives room for longer electrical trailing edges (Fig. 3a).

After having enlarged the amorphization window by a modification of the thermal environment of the antimony, the devices with a 40 nm thin $SiO_2$ heat barrier are suited to experiments at higher ambient temperatures. A higher base temperature has two major impacts on the ability to amorphize, similar to the influence of the heat barrier: on the one hand the same electrical heating power will melt a larger volume of antimony; on the other hand the process of cooling from the melt will be slowed down due to the smaller temperature gradients resulting in a longer time spent in the temperature regime of fast crystallization. Again, as might be expected for a material with a very strong tendency to rapidly crystallize, expanded recrystallization dominates over a somewhat larger molten volume to begin the quenching with. Overall, the amorphization window shrinks for higher ambient temperature, but it remains clearly open up to 250 K (Fig. 3b). Furthermore, in the measurement series performed at 100 K, 150 K, 200 K and 250 K there is no indication for a sudden end of the feasibility to amorphize Sb even above room temperature.

However, when turning towards higher ambient temperatures, quenching the melt rapidly enough to pass through the high-temperature regime without substantial crystallization is only one of the challenges posed. At higher base temperatures the stability of a successfully created amorphous state weakens considerably. In order to quantify this stability, we observe how the device resistance evolves and determine the time that has passed since the electrical melting pulse until the device resistance is only a factor of two above its crystalline value. This crystallization time appears to depend on temperature in an Arrhenius way, at least over the four orders of magnitude in time covered by our experiments (Fig. 4 and supplementary information section 'Arrhenius behaviour of crystallization time'). For the devices with 5 nm thick antimony investigated until now, the 'retention time' above room temperature is in the range of few seconds and below, still far from the stabilities one is used to from established phase change materials. However, narrowing the confinement of a glass between interfaces with neighbouring materials can be an effective way to stabilize it by restricting its structural dynamics[26-29]. Therefore, we investigated how the robustness of antimony against crystallization can be influenced by scaling down its thickness. Indeed, the stability is boosted by more than 100 K in base temperature or by many orders of magnitude in time respectively, when reducing the thickness of Sb from 10 nm to 3 nm (Fig. 4).

The retention time we have achieved already meets the requirements for some computing-in-memory tasks[30] and is orders of magnitude higher than that of dynamic random access memory (see supplementary information section 'Retention times in different memory applications'). For enhanced stability of the amorphous state, besides a further well-controlled reduction of the thickness, systematic investigations of alternative neighbouring materials with particular focus on their atomic-scale roughness[31] and rigidity will be instrumental in enhancing amorphous stability. Ultimately, when moving towards next-generation memory devices with three-dimensional nanoscale confinement, also the effects of mechanical stress must be expected to play an important role for the stability of the amorphous states. Quantitative analyses in theory, in simulation and in experiments studying these dependencies under careful control of all mentioned aspects are lacking. Our work demonstrates what impact such research could have for realizing reliable phase-change based devices integrated with highest spatial density.

With this work we advocate a paradigm shift for the research on phase change materials for information processing. Instead of following the wisdom of the past and proposing ever new mixtures with questionable chances of being achievable and maintainable in ultra-scaled structures, we turn towards the most radical simplification on the material side. As a consequence, in this context, discussions about how a particular composition might be necessary for achieving improved phase-change functionality become obsolete. In contrast, quantitative knowledge of effects related to nanoscale confinement emerges as a matter of highest importance.

**FIGURES**

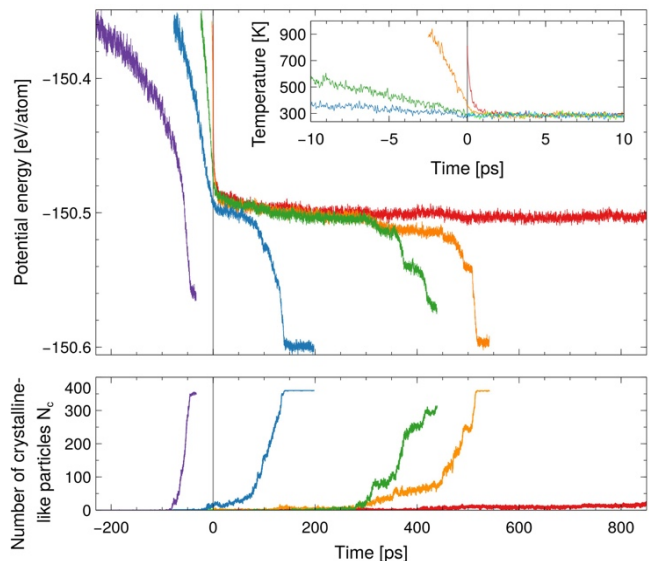

**FIG. 1: Increase of quenching rate hinders crystallization.** An Ab-initio Molecular Dynamics model of 360 Sb atoms is quenched from the melt with varied rates (3 K/ps purple, 9.5 K/ps blue, 30 K/ps green, 300 K/ps orange, abrupt quenching in a single simulation step red) reaching its target temperature of 300 K at time t=0 ps (see inset). A drop in potential energy (top panel) below around -150.5 eV per atom accompanied by an increase of the number of crystalline-like particles $N_c$ (bottom panel) marks crystallization. Details about the definition of $N_c$ is given in the supplementary information section 'AIMD simulations of quenching rate variations'. The simulation with the slowest quenching rate (purple) was aborted before reaching the target temperature as crystallization started already during the quenching period.

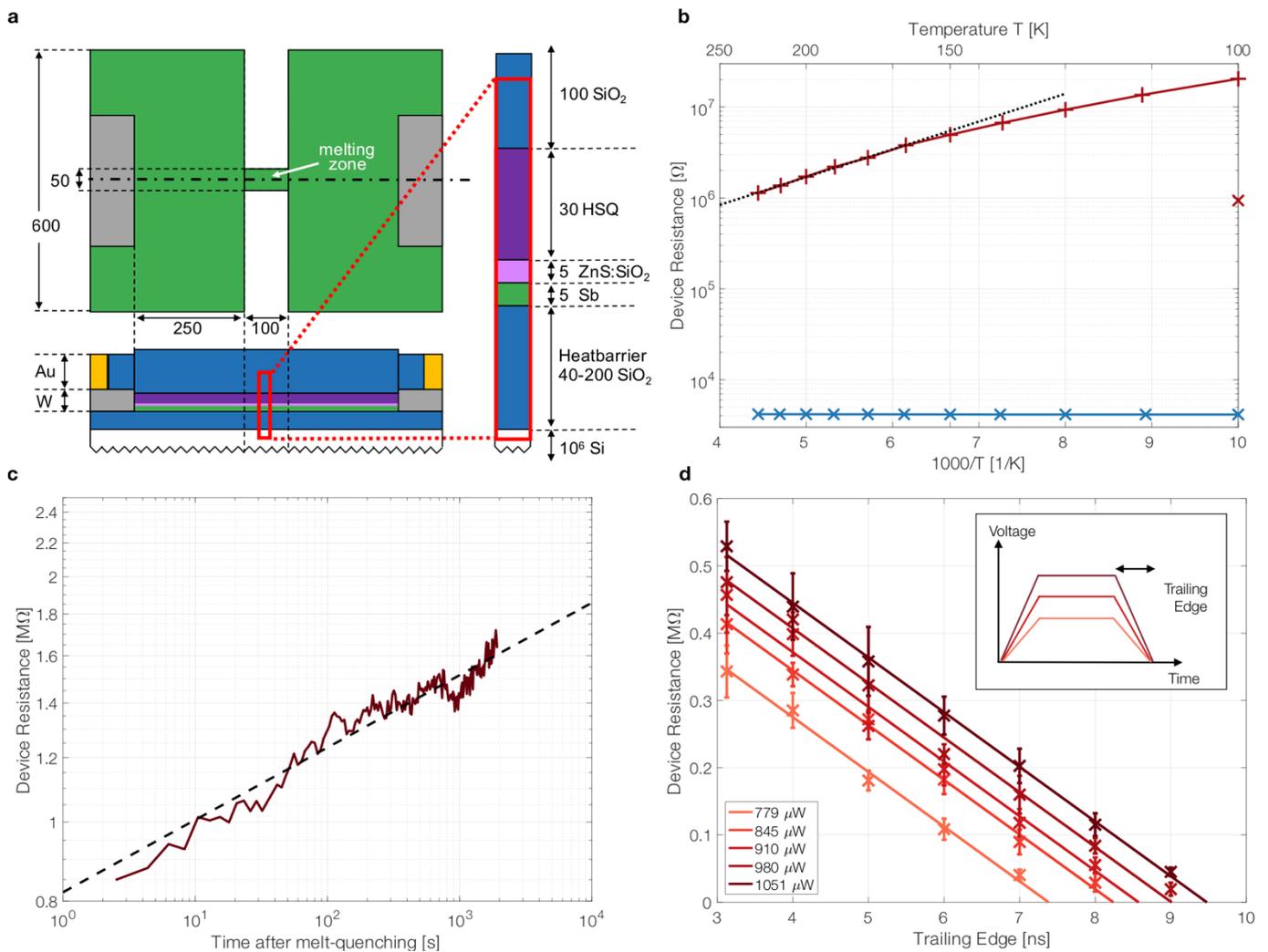

**FIG. 2: Creation of melt-quenched amorphous Sb in electrical switching experiments.** (a) Device geometry with dimensions given in nm. (b) Device resistance in a fully crystalline device (blue x) and directly after successful amorphization at ambient temperature of 100 K with electrical pulse power of 957 μW (red x). Resistance series (red +) taken during stepwise temperature decrease after annealing at 225 K reveals thermal activation of the electrical transport with an apparent activation energy of 0.065 eV between 225 K and 175 K (dotted black line). (c) Temporal increase of device resistance after melt-quenching at 100 K characteristic for amorphous phase change materials. Dashed line is a fit of the power law conventionally used to describe such a resistance drift. (d) Influence of trailing edge of applied electrical pulses on resulting device resistance at a base temperature of 100 K. Different shades of red represent pulse powers between 779 and 1051 μW. Variation of pulse parameters are illustrated in the inset to panel (d). The error bars denote the standard deviation determined from five identical electrical excitations. Each amorphization (b-d) was induced by a trapezoidal voltage pulse with 50 ns plateau duration. The trailing edge in (b) and (c) was 3 ns.

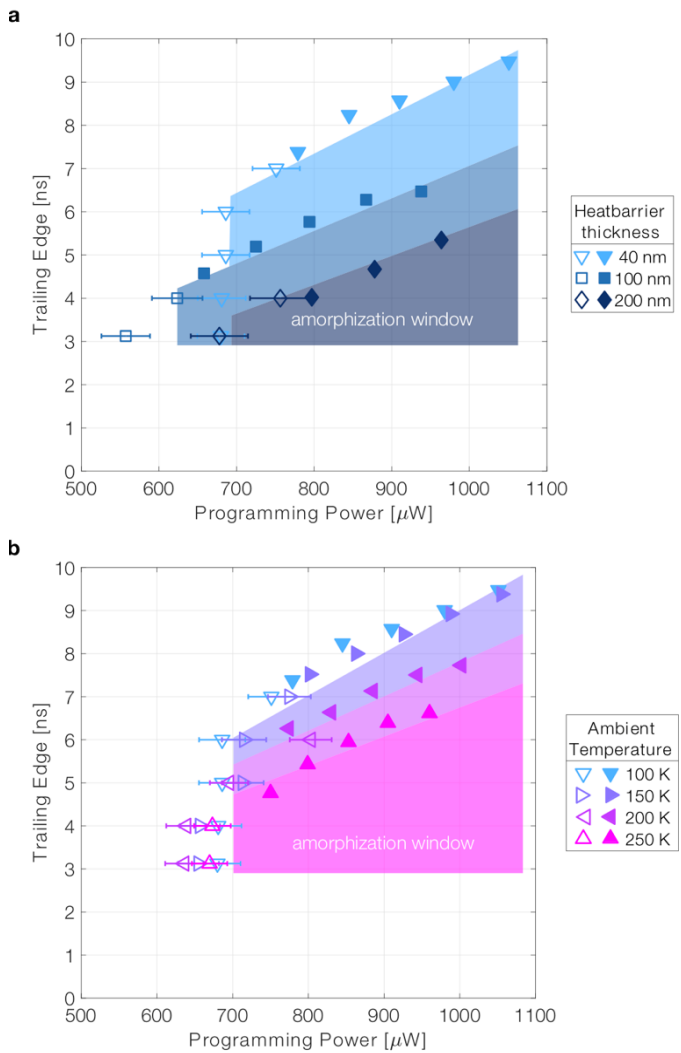

**FIG. 3: Controlling the amorphization window.** (a) Increasing heat dissipation into the thermal environment of the PCM device by thinning the heat barrier enlarges the amorphization window enough to keep it open even when (b) the ambient temperature is increased towards room temperature. Each data point, in (a) and (b), is based on a series of trailing edge experiments as plotted in Fig. 2d. Filled symbols represent the longest trailing edge at a certain pulse power that still results in a device resistance higher than the crystalline state (forming an upper bound of the amorphization window); empty symbols mark the minimum programming power below which no significant resistance increase was achievable with a certain trailing edge (forming the left border of the amorphization window). More accurately, the minimum programming power lies between the highest programming power that did not yet increase the device resistance above its crystalline starting point on the one hand side and the lowest programming power that did significantly raise the device resistance on the other hand side. Those empty symbols are positioned in the middle between those two power values. The error bars reach from the first to the latter power value. The base temperature in (a) is 100 K. The heat barrier thickness in (b) is 40 nm. The thickness of Sb is 5 nm in (a) and (b). Because the amorphization window does not change considerably going from 100 K to 150 K (b), it is coloured only once (purple). The successfully created amorphous states are stable for much longer than the few seconds it takes to measure their resistance after the melting pulse.

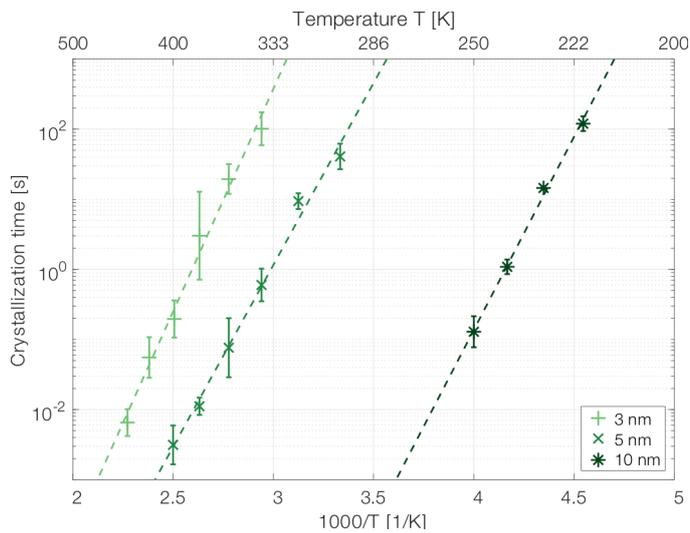

**FIG. 4: Improving robustness against crystallization by narrowing the confinement of the elemental glass.** While reducing the device thickness from 10 nm (dark green asterisk) to 5 nm (full green x) the increase in absolute crystallization times is already very strong, further confinement from 5 nm to 3 nm (light green +) even results in a steeper temperature dependence of the crystallization time. To obtain an average value for the crystallization time at a given temperature, devices were melt-quenched (and recrystallized) at least five times. The error bars correspond to the standard deviation of $\log_{10}$(crystallization time).

**Supplementary Information** is available in the online version of the paper.

**Acknowledgements** The research leading to these results has received funding from the People Programme (Marie Curie Actions) of the European Union's Seventh Framework Programme FP7/2007-2013/ under REA grant agreement No 610781, from the European Research Council (ERC) under the European Union's Horizon 2020 research and innovation programme (grant agreement Numbers 640003 and 682675), and from Deutsche Forschungsgemeinschaft (DFG) through the collaborative research centre Nanoswitches (SFB 917). We also acknowledge the computational resources provided by JARA-HPC from RWTH Aachen University under Projects No. JARA0150 and JARA0176. Finally, we thank Evangelos Eleftheriou and Wabe W. Koelmans at IBM Research Zurich for his support of this work.

**Author Contributions** B.K., A.S. and M.S. conceived and designed the experiments; V.P.J. and X.T.V. fabricated the Sb-based devices with support from I.G.; O.C.-M. analysed the integrity of the deposited Sb via Atom-Probe Tomography; B.K. performed the experiments supported by M.L. and advised by M.S. and A.S.; B.K. and M.S. analysed the data; I.R. performed the computer simulations and analysed the data, with help from R.M.; M.S. wrote the manuscript with input from all the authors.

**Author Information** Reprints and permissions information is available at www.nature.com/reprints. The authors declare no competing financial interests. Readers are welcome to comment on the online version of the paper. Correspondence and requests for materials should be addressed to M.S. (martin.salinga@rwth-aachen.de) and A.S. (ase@zurich.ibm.com).

## Methods

**AIMD:**

Different from a very recent publication simulating supercooled liquid Sb[32], for our Ab initio Molecular Dynamics (AIMD) simulations based on Density Functional Theory (DFT) we employed the second generation Car-Parrinello scheme[33] which is implemented in the Quickstep code of the CP2K simulation package[34] (a more detailed discussion of our AIMD simulations is given in the supplementary information sections 'AIMD methodology' and 'AIMD simulations of quenching rate variations'). Effects due to the finite size of the used simulation box (360 atoms) can lead to a significantly reduced crystallization time (see the effect of increasing the system size to 540 and even 720 atoms in the supplementary information section 'Finite size effects in AIMD simulations', in particular Fig. S4). Also, the speed of the crystallization process is very sensitive to the atomic density, which is quite difficult to determine accurately for experimentally realistic volumes. As an example: allowing the Sb atoms to fill a volume that is enlarged by only 7 % reduces the stress by a factor of 5, leading to a 10 times higher stability against crystallization at 450 K (a detailed discussion is provided in the supplementary information section 'AIMD simulations with different densities').

**Device fabrication:**

The devices were fabricated on a silicon substrate with a thermally grown $SiO_2$ top layer of varying thickness (40, 100, 200 nm) acting as thermal and electrical insulation. The Sb with a thickness of 3, 5 or 10 nm and a 5 nm thick capping layer of $(ZnS)_{80}(SiO_2)_{20}$ were deposited by sputtering with an average rate of <0.1 nm/s. The purity of the deposited Sb was assured to be >99.9 % using Atom-Probe Tomography. The lateral device shape depicted in Fig. 2a was then realized using using e-beam lithography with hydrogen silsesquioxane (HSQ) resist and ion milling. The patterned structure was immediately passivated with additional ~18 nm of sputtered $SiO_2$. In order to electrically contact the Sb layer, holes were ion-milled into the capping layers in another e-beam lithography step, ensuring that the etching process is stopped once it reached the SiO₂ heat barrier underneath the Sb. Then, a third e-beam lithography step with lift-off was used to shape a sputter-deposited layer of tungsten into lateral electrical leads connected to the phase change material. Also, for better switching stability, a titanium resistor (ranging between 2 to 4 kOhm) was added in series with the Sb device using lift-off. The whole chip was then encapsulated with a layer of 80-nm-thick sputtered SiO₂. Finally, e-beam lithography and reactive-ion etching were employed to locally open the encapsulation before adding gold probe pads (200 nm, sputter-deposited, and shaped via optical lithography and lift-off).

**Electrical testing at cryogenic temperatures:**

The electrical measurements were performed in a liquid-nitrogen-cooled cryogenic probing station (JANIS ST-500-2-UHT) operating between 77 to 400 K. The temperature was controlled using two heaters with powers of 50 and 25 W, calibrated Lakeshore Si DT-670B-CU-HT diodes with an accuracy of <0.5 K at four positions in the chamber and a Lakeshore 336 Automatic Temperature Controller. A radiation shield is fixed above the sample mount and thermally connected to the nitrogen out flux isolates. The pressure inside the chamber is reduced below $10^{-5}$ mbar to avoid heat exchange via convection and water condensation at low temperatures. A high- frequency Cascade Microtech Dual-Z probe is used to contact the devices inside the cryogenic probe station. It is thermally connected via cooling braids to the bulk metallic sample holder keeping sample and probe at the same temperature.

A Keithley 2400 Source Measure Unit was used for DC measurements of the device resistance (at a constant voltage of 0.1 V). An Agilent 81150A Pulse Function Arbitrary Generator sent the voltage pulses to the device under test and a Tektronix oscilloscope (TDS3054B/DPO5104B) recorded applied voltage and transmitted current signals. Mechanical relays (OMRON G6Z-1F-A) were used to switch between AC and DC configuration. All switching voltage pulses had a trapezoidal shape with a plateau length of 50 ns and equal leading and trailing edge. The pulse power experienced by the Sb device is calculated based on the plateau values of the time-resolved current- and voltage traces recorded by the oscilloscope under consideration of the series resistor. In order to exclude gradual, irreversible changes in the tested device being responsible for the observed effect of the varying trailing edges (Fig. 2d), the first measurement series (with the shortest trailing edge) was always reproduced at the end of all other experiments with longer trailing edges.

**Data availability:**

The data that support the findings of this study are available from the corresponding author upon reasonable request.

**References cited only in the Methods section:**

**Title of the main article:** Monatomic phase change memory

**Authors:** Martin Salinga*[1,2], Benedikt Kersting[1,2], Ider Ronneberger[1,2], Vara Prasad Jonnalagadda[1], Xuan Thang Vu[2], Manuel Le Gallo[1], Iason Giannopoulos[1], Oana Cojocaru-Mirédin[2], Riccardo Mazzarello[2], Abu Sebastian[1]

[1] IBM Research-Zurich, CH-8803 Rüschlikon, Switzerland

[2] RWTH Aachen University, D-52074 Aachen, Germany

**Contents:**

# AIMD methodology

For our Ab initio Molecular Dynamics (AIMD) simulations based on Density Functional Theory (DFT) we employed the second generation Car-Parrinello scheme[1] which is implemented in the Quickstep code of the CP2K simulation package[2]. In this scheme the efficiency of the Car-Parrinello method is combined with the large time steps used in Born-Oppenheimer Molecular Dynamics (BOMD). Instead of solving the Kohn-Sham equations self-consistently at each molecular dynamics step or using the Car-Parrinello equations of motion, a predictor-corrector algorithm, more specifically the always stable predictor-corrector method by Kolafa[3], is used to propagate the expansion coefficients of the wave function basis set. Using this scheme it is possible to maintain the system close to the Born-Oppenheimer surface without the need of using small time steps. As a consequence, a computational speedup of approximately a factor 20 to 30 with respect to the standard (BOMD) are typically achieved. However, the dynamics of this method is dissipative with some intrinsic friction coefficient of the system and thus cannot be used in the microcanonical ensemble (NVE). However, it can be used in the canonical ensemble (NVT). The implementation in CP2K utilizes a stochastic Langevin thermostat to control the intrinsic dissipation. This method has been proven to be to provide reliable results for many systems, for example in liquid $H_2O$, Si, l-$SiO_2$[1,4] and in particular it was successfully used for simulations of typical phase-change materials GeTe[5], $Ge_2Sb_2Te_5$[6], $Ag_4In_3Sb_{67}Te_{26}$[7].

In the present simulation, we use this scheme for two types of systems, namely pure Sb and an interfacial model of Sb/$SiO_2$. For the latter system, it was necessary to increase the number of predictor-corrector steps to properly thermalize the two subsystems (Sb and $SiO_2$). Generalized gradient approximation (GGA) to the exchange-correlation function[8] and scalar-relativistic Goedecker pseudopotentials[9] (PP) are used for all the simulations. Triplezeta plus polarization Gaussian-type basis set are employed for the expansion of the Kohn-Sham orbitals and the charge density is expanded in plane waves with a cut-off of 300 Ry. Periodic boundary conditions are applied and the Brillouin zone is sampled at the $\Gamma$ of the supercell.

# AIMD simulations of quenching rate variations

For the generation of stable supercooled liquid and amorphous states of phase-change materials very large quenching rates are required to avoid crystallization. Due to limitations of the computational time scales, the quenching rate employed in simulations of typical phase-change materials[6,7,10-12] are on the order of $10^{13}$ K/s which is around two orders of magnitude above typical experimentally realized values. Nevertheless, a variation of the quenching rates in the simulations on a computationally affordable time scale might give qualitative statements on the stability of supercooled and amorphous phase-change materials. In the present set of simulations, we performed AIMD simulations of pure Sb by changing the quenching rate $\gamma$ over two orders of magnitude. For that we considered a model containing 360 atoms in an orthorhombic supercell of 21.77 x 22.62 x 22.78 Å$^3$ which corresponds to the density of liquid Sb and is equal to the experimental value of 6.49 g/cm$^3$ (Ref. [13]) at T=915 K. Five different simulations with quenching rates $\gamma$ = 3, 9.5, 30, 300 K/ps and an idealized maximum quenching rate of "$\infty$ K/ps" which is obtained by directly setting the target temperature from ~930 K to ~300 K (see Fig. S1). Our stochastic Langevin thermostat achieves the temperature within one picosecond which corresponds to ~630 K/ps. The quenching rates are realized by changing the target temperature in a stepwise fashion, e.g. $\gamma$ = 30 K/ps was achieved by decreasing the temperature by 15 K every 0.5 ps. After the quenching, the models were annealed at room temperature.

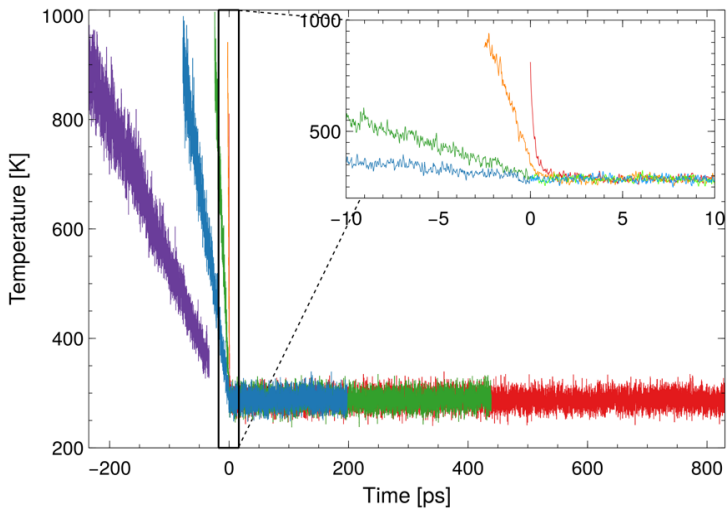

**Fig. S1:** The temperature profile of the models with different quenching rates, ranging γ = 3 K/ps to ∞ K/ps. The inset shows the zoomed-in region of fast quenching which is indicated by a rectangle.

The results of this set of simulations are summarized in Fig. 1 of the main text. It shows the time evolution of the potential energy (dark colours) and the number of crystalline-like atoms $N_C$ (light colours) for all the models.

The $N_C$ is determined based on a bond order correlation parameter, a definition of which is given in Ref. [14], Chapter 3, pages 60-63. We distinguish crystalline-like atoms by computing the bond order correlation parameter $q_4^{dot}(i)$ for each atom i and using a threshold value of 0.65. Liquid-like and amorphous-like atoms display values close to 0, because the bond orientations between neighbouring atoms are uncorrelated, whereas crystalline-like atoms take values close to 1. The statistical distribution of the $q_4^{dot}$ parameter for different phases of Sb are shown in Fig. S2. The distributions in the liquid and amorphous phases show negligibly small overlap with the distributions of perfectly crystalline and recrystallized phase at values close 0.65. We choose this value as a threshold and define an atom to be crystalline-like if its $q_4^{dot}$ exceeds that threshold.

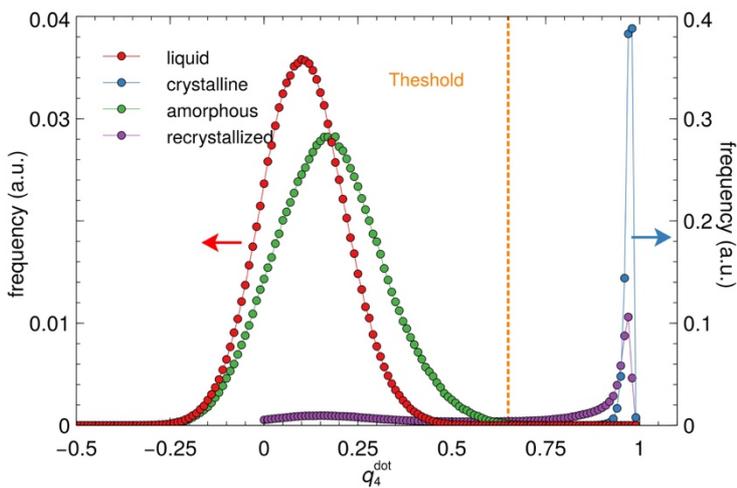

**Fig. S2:** The distributions of $q_4^{dot}$ for liquid (red), amorphous (green), crystalline (blue) and recrystallized (purple) phases of Sb. For the corresponding computation of liquid, amorphous and recrystallized phases, segments from the trajectories of 360-atom models in the corresponding phases were taken. For the computation of crystalline phase, a short run of a perfectly crystalline model at 600 K was performed. Since the distribution in the crystalline phases are narrow, a second axis (indicated by the blue arrow) with a larger scale is used for the crystalline and recrystallized distributions. The threshold value used to define crystalline-like atoms is indicated as a dashed vertical line.

The model with the slowest quench rate of $\gamma = 3$ K/ps crystallizes already during the quenching process. This result is consistent with the observation of very short crystallization times in a temperature range around ~500 to 600 K. Prior to the simulations discussed here, various models of supercooled Sb (360 atoms) were annealed at different temperatures ranging from 700 K to 300 K. Detailed discussion of the results for that set of simulations can be found in Ref. [14] (Chapter 7). All the models in the temperature range of 400-550 K crystallized very quickly whereas in the temperature range above ~600 K and below ~300 K no fast crystallization occurred.

The remaining four models of our present study reached room temperature without crystallization. The model with $\gamma = 9.5$ K/ps contains a small fraction of crystalline-like atoms ($N_C$~10) at the time when the target temperature was set to 300 K (t = 0) which steadily increases until the onset of crystallization. $N_C$ increases more slowly in the models with $\gamma = 30$ K/ps and $\gamma = 300$ K/ps. The onset of crystallization is at ~300 ps in both models. However, the subsequent crystallization of the faster quenched model (with $\gamma = 300$ K/ps) occurs notably slower resulting in a longer crystallization time. The model with the highest quenching rate $\gamma = \infty$ K/ps. sets itself apart from all the other models in that it does not crystallize even after ~800 ps. Nonetheless, the number of crystalline-like atoms $N_C$ in this model grows slowly over time suggesting that it will eventually crystallize once it reaches a critical size.

In our models we observe a consistent trend of increased stability of amorphous Sb with faster quenching rates. This result suggests that qualitatively employing higher quenching rates can lead to a more stable amorphous phase of Sb. Nevertheless, one should keep in mind that the crystallization is a stochastic process with varying crystallization times at a given temperature. A more quantitative analysis would require a larger number of independent simulations to improve the statistics of the crystallization times. This is, however, currently neither feasible with AIMD nor needed for the present study due to other limitations of the model (see sections below).

## AIMD simulations with different densities

A parameter that is known to have an influence on the stability of the amorphous state is the density. In the current study we used the liquid density of $\rho_{high} = 6.49$ g/cm³. Here, we address the question on how the density affects the stability of the supercooled and amorphous phase by generating an additional model with 7 % lower density of $\rho_{low} = 6.03$ g/cm³ by quenching with $\gamma = 30$ K/ps. We chose a model with $\rho_{high}$ from the previous simulations at 500 K which falls into the range of fast crystallization and performed the same annealing run for the new model with $\rho_{low}$ at that temperature. The results of these two models are compared in Fig. S3. As can be seen from both the potential energy and the number of crystalline-like atoms $N_C$, the onset of crystallization for the model at lower density occurs at a significantly later point in time.

For both models temperature and virial stress values were obtained averaging values in the first 20 ps of the simulation. For the virial stress, the average of the diagonal components $\langle\sigma_{ii}\rangle = (\sigma_{xx} + \sigma_{yy} + \sigma_{zz})/3$ resulted in $\langle\sigma_{ii}\rangle = 0.28 \pm 0.19\ GPa$ at a temperature of $T = 503 \pm 23\ K$ for the $\rho_{low}$ model and $\langle\sigma_{ii}\rangle = 1.40 \pm 0.18\ GPa$ at $T = 509 \pm 20\ K$ for the $\rho_{high}$ model. While the $\rho_{low}$ model is almost stress-free displaying values close to zero in the stress tensor, the average of the virial stress components of the high density model is five times larger than that of the low density model. This result suggests that supercooled and amorphous phase of Sb are destabilized by application of compressive stress. A similar finding was provided in Ref. [15] in which the stability of amorphous Ge-doped Sb (i.e. Ge$_7$Sb$_{93}$ and Ge$_6$Sb$_{94}$), closely related phase-change compounds, was shown to decrease experimentally upon compressive stress.

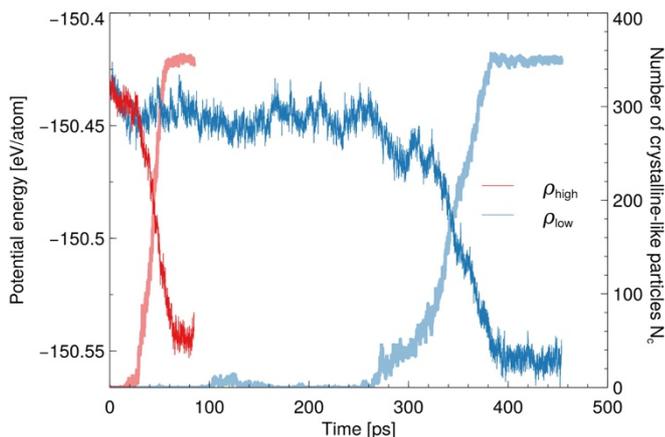

**Fig. S3:** The potential energy U (dark) and the number of crystalline-like particles $N_c$ (light) of two supercooled models at T=500 K with two different densities: $\rho_{high}$ = 6.49 g/cm$^3$ (red) and $\rho_{low}$ = 6.03 g/cm$^3$ (blue). Both models are obtained with the same quenching rate of $\gamma$ = 30 K/ps.

## Finite size effects in AIMD simulations

Finite size effects are inevitable in the typical system sizes employed in AIMD simulations. Because of the periodic boundary conditions atoms interact with their periodic images in a spurious way. In Ref. [14], we showed that the free energy barrier for nucleation in very small systems (on the order of 50 to 70 atoms) can take extremely small values. Simulations of such small systems can be inflicted with severe artefacts and should be therefore avoided. On the other hand, the system size is the main factor which determines the computational cost in AIMD simulations. Our system size of 360 atoms lies between the typical system sizes of 216 (GeTe) and 460 (Ge$_2$Sb$_2$Te$_5$) atoms used in several AIMD studies [12,14,16] of phase-change materials. Currently the largest model used in AIMD simulations of crystallization contains 900 atoms (Ge$_2$Sb$_2$Te$_5$)[14].

Classical molecular dynamics simulations are computationally much cheaper than AIMD. Thus finite system size effects can be easily reduced in such simulations by increasing the model size. However, classical simulations require reliable interatomic potentials which must be able to accurately describe chemical bonding in Sb. Currently no such potential for Sb is available. Therefore, we restrict ourselves to AIMD simulations of models with affordable sizes. We qualitatively assess the system size effects by employing two larger models of Sb ($\rho_{high}$ = 6.49 g/cm$^3$) containing 540 and 720 atoms inside a cubic cell with dimensions 25.62 x 25.62 x 25.62 Å$^3$ and 28.20 x 28.20 x 28.20 Å$^3$, respectively. Supercooled models for these system sizes were obtained by quenching them with $\gamma$ = 30 K/ps. Each of these three models was annealed at T = 450 K resulting in the data plotted in Fig. S4. The crystallization time increases with model size. This is consistent with our argument about the reduction of the nucleation barrier with decreasing model sizes.

Despite the difference in the nucleation barriers we expect qualitatively similar behaviour for the larger models regarding quenching rate and density effects, albeit on longer times scales. We would like to stress that the observation of crystallization itself in the larger models are very important, since this shows that fast crystallization in the smaller models are not merely artefacts of the model size.

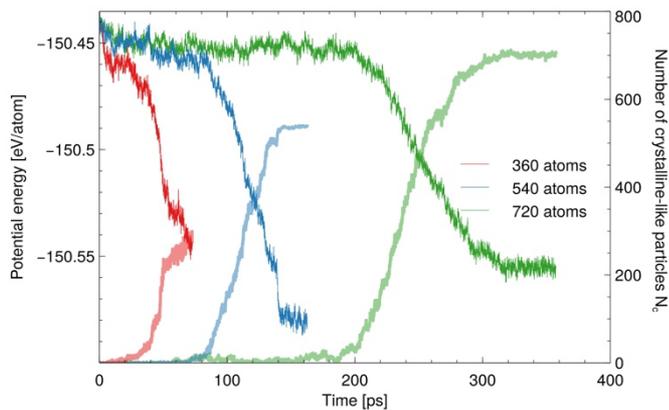

**Fig. S4:** The potential energy U (dark) and the number of crystalline-like particles Nc (light) of supercooled Sb models at T=450 K with different system sizes of 360 (red), 540 (blue) and 720 atoms (green). All the models are obtained with the same quenching rate of $\gamma$ = 30 K/ps.

## Simulation of Sb/SiO2 interface at high temperatures

A question relevant for the interpretation of our experimental data is whether we must expect mixing of the confining dielectric material with the phase-change material (Sb). The likelihood for such a mixing would certainly be highest during the most intense reset operation of the device involving melting of the Sb. We addressed this by running simulations of an interface model at high temperatures well above the melting temperature
of Sb ($T_{m,Sb}$ = 903 K). For this purpose, we constructed an interface model which is composed of a 360-atom model of liquid Sb and a separately generated model of amorphous SiO$_2$ containing 288 atoms. Two different models with two different densities were considered. Four different simulation temperatures were chosen, namely, 1000 K, 2000 K, 3000 K and 4000 K. We find no mixing of atoms at 1000 K and 2000 K in the respective simulation times ranging between 35 ps to 55 ps. Only at temperatures above the melting point of SiO$_2$, $T_{m,SiO2}$ ≈ 2000 K, (i.e. at 3000 K and 4000 K) the different atomic species start to mix with each other. At temperatures well above $T_{m,SiO2}$, the Si and O atoms show very high atomic mobility (diffusion) indicative of SiO$_2$ becoming liquid. Based on these results we can exclude atomic mixing of Sb with the neighbouring dielectric material for temperatures up to 2000 K.

## Observation of threshold switching

Besides the temperature dependence of the resistance (Fig. 2b) and the temporal evolution of the resistance (Fig. 2c) there is another characteristic attribute of amorphous phase change materials observable in our melt-quenched Sb devices: The sudden breakdown of resistance when the applied voltage reaches certain threshold (see exemplarily Fig. S5). Also the linear dependence between threshold voltage and device resistance is agreement with what has been observed for traditional phase change memory devices in the past. The device resistance can approximately be assumed to be a measure for the size of the amorphous mark along the length of the Sb line.

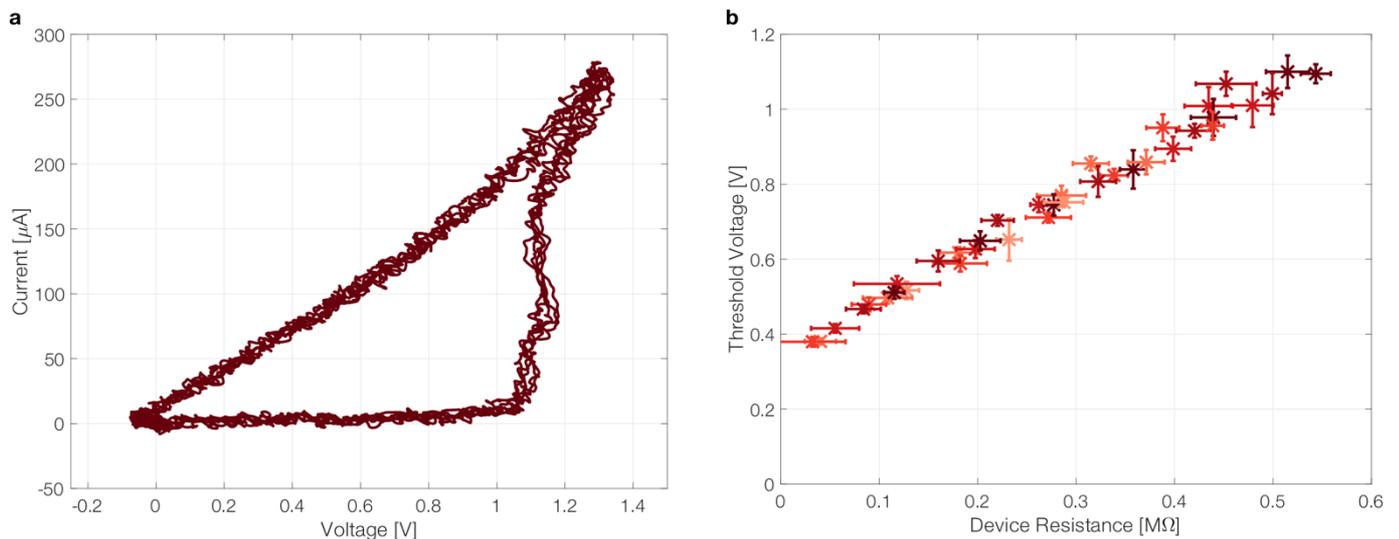

**Fig. S5: Current response to voltage applied to an amorphized Sb device shows threshold switching.** Device geometry: 5 nm thick Sb on 40 nm thick SiO$_2$ heat barrier; ambient temperature: 100 K; electrical pulse inducing amorphization: 50 ns plateau, 3 ns leading and trailing edge, 1051 µW pulse power; voltage sweep testing the threshold behaviour: triangular pulse with 200 ns leading and trailing edge. The measurement was repeated five times. (a) Each of the resulting current-voltage characteristics is plotted demonstrating remarkable reproducibility. (b) Threshold voltage (defined as voltage where the current response surpasses 30 µA) as a function of device resistance. The colour code is according to the one given in Fig. 2d of the main text, i.e. representing various pulse powers of the amorphization pulse. The error bars denote the standard deviation determined from five identical electrical excitations.

## Arrhenius behaviour of crystallization time

Fig. 4 of the main text shows that the temperature dependence of the crystallization time can be described reasonable well by an Arrhenius behaviour. The according activation energies resulting from the fits represented by the dashed lines in Fig. 4 are 1.09 ± 0.19 eV for 10 nm Sb, 1.03 ± 0.13 eV for 5 nm Sb, and 1.26 ± 0.20 eV for 3 nm Sb.

The effect of uncontrolled impurities should be included in the margin of error of those measurement series (in Fig. 4), since experiments on several different devices have contributed to the crystallization data for each thickness.

## Definition of an amorphization window

The concept of an amorphization window in the parameter space of power and trailing edge length of the electrical amorphization pulses is discussed in the main text. The definitions of the limits of such an amorphization window are given in the figure caption of Fig. 3 of the main text. Here (with Fig. S6), an illustration is added as a visual assistance for an easier understanding of how each limiting point in Fig. 3 is derived from a whole series of electrical measurements.

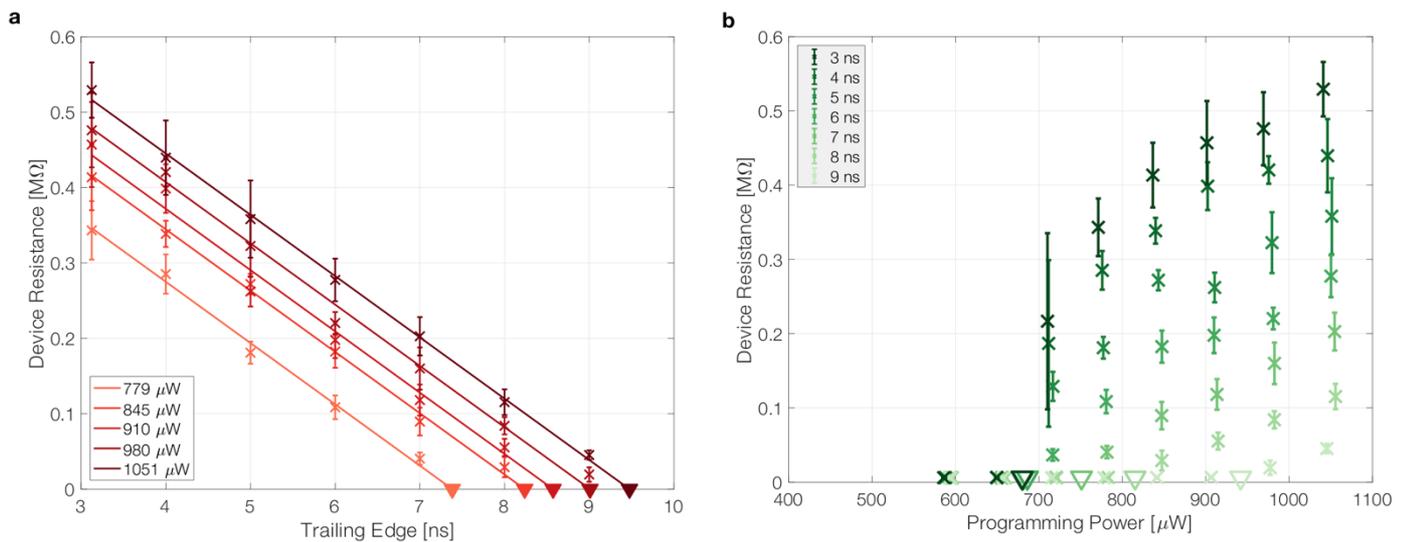

**Fig. S6: Illustration of the derivation of the limits of the amorphization windows from series of amorphization experiments.**

(a) Device resistance resulting from an electrical excitation with a certain trailing edge length (horizontal axis) and pulse power (shades of red). The error bars denote the standard deviation determined from five identical electrical excitations. Filled triangles represent the longest trailing edge at a certain pulse power that still results in a device resistance higher than the crystalline state. More accurately, the limiting trailing edge value for each pulse power is read off where the extrapolation of the apparently linear decrease of the device resistance with trailing edge length hits the resistance value of the fully crystalline device (practically zero on a linear MΩ scale). These values of a trailing edge length and according pulse power form the upper bound of an amorphization windows in Fig. 3.

(b) Device resistance resulting from an electrical excitation with a certain trailing edge length (shades of green) and pulse power (horizontal axis). The error bars denote the standard deviation determined from five identical electrical excitations. Empty triangles mark the minimum programming power below which no significant resistance increase was achievable with a certain trailing edge. More accurately, the minimum programming power lies between the highest programming power that did not yet increase the device resistance above its crystalline starting point on the one hand side and the lowest programming power that does significantly raise the device resistance on the other hand side. Those empty triangles are positioned in the middle between those two power values. The error bars of those data points marking the left border of the amorphization window in Fig. 3 reach from the first to the latter power value.

## Critical cooling rate for glass formation

In the AIMD simulations shown in Fig. 1 crystallization took place during the cooling phase when the quenching rate was set to 3K/ps. Already with a quenching rate of 9.5K/ps there is no sign of crystallization during the cooling phase anymore. As we discuss in the methods section on "AIMD" and in the according paragraphs in the supplement ("AIMD simulations with different densities" and "Finite size effects in AIMD simulations") one has to expect that in reality the timescales for crystallization are significantly longer. That is why we prefer to only take qualitative trends from our simulations.

When attempting to determine a critical cooling rate in case of the experiments there is another problem. In a self-heated nanoscale device, temperature across the material under test is very inhomogeneous. Also, the temporal evolution of the temperature during cooling is not simply linear. Thus, there is not one cooling rate the material under test experienced. So even if one determines the temporal evolution of the temperature for every part of the molten material by simulating the device structure (e.g. via Finite Elements Method), which is highly unreliable anyways due to the lack of many significant input parameters (like thermal interface resistances) as a function of temperature, one will not be able to condense the outcome into a simple critical cooling rate.

That is why we decided to modify the cooling of the material from the melt by using a quantity which we can control very well, i.e. the trailing edge of our electrical excitation. Evidently, it is possible to significantly influence the cooling times through a variation of the trailing edges in the range of few nanoseconds. Therefore, the cooling times themselves must be on a similar time scale. Thus, one can estimate that the temperature in the material was reduced by several hundred Kelvin (from above melting temperature to base temperature) in few nanoseconds. Under these conditions glass formation of pure Sb is apparently possible.

## Retention times in different memory applications

In the application field of Storage Class Memories (the one currently addressed by Intel's and Micron's 3D-Xpoint technology) alone there can be different flavours: from storage-mapped (longer retention, but not as fast) to memory-mapped (faster and higher endurance, but allowed to be much more volatile).

P. Cappelletti (Micron) writes in his 2015 IEDM article entitled „Non volatile memory evolution and revolution"[17]: "While present visibility gives emerging memories little chance to compete with NAND in cost and density, there is definitely a great opportunity for emerging memory to help close the constantly increasing performance gap between DRAM and NAND. However, even that space, often referred to as the Storage Class Memory space, is too wide to be covered by a single technology; it is more practical to consider two flavors of SCMs, associated to two different areas of applications: memory mapped and storage mapped (fig.8). The memory mapped SCMs (…) do not need to be "true" non-volatile (a retention time much longer than DRAM refresh time is sufficient)." In that fig. 8 the non-volatility criterion is specified for memory-mapped SCM as longer than ten times the DRAM refresh interval, which is currently 64 ms. So, the necessary retention time for such a memory application is less than a second.

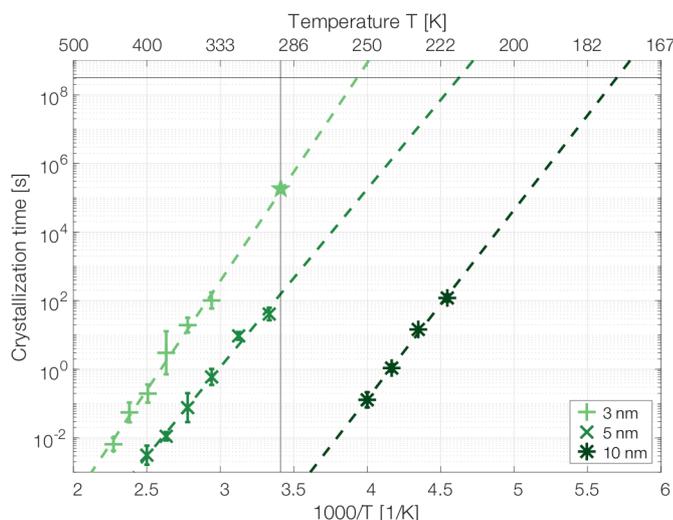

**Fig. S7: Extrapolation of crystallization data towards long times:** The vertical line marks room temperature (20°C), the horizontal line marks 10 years. Besides the experimental data depicted and fitted in Fig. 4, here we added the outcome of a single measurement at room temperature (50.8 hours) confirming the extrapolation of the data taken at elevated temperatures.

Beyond Storage Class Memories, there are other currently emerging applications of phase change memory devices. Given the explosive growth in data-centric cognitive computing and the imminent end of CMOS scaling laws, it is becoming increasingly clear that we need to transition to non-von Neumann computing architectures. In-memory computing and brain-inspired neuromorphic computing are two approaches that are being actively researched.

In in-memory computing, the physical attributes and dynamics of memory devices are exploited to perform certain computational tasks in place. For example, PCM devices organized in a crossbar array can be used to solve linear equations, performing part of the computation in an iterative solver. In one such mixed-precision approach, depicted in Fig. S8, crossbar arrays of PCM devices are used to perform inexact analog matrix-vector multiplications using Ohm's and Kirchhoff's circuit laws in an iterative Krylov-subspace solver. A high-precision computing unit is used to iteratively improve the solution accuracy until a desired tolerance is reached.

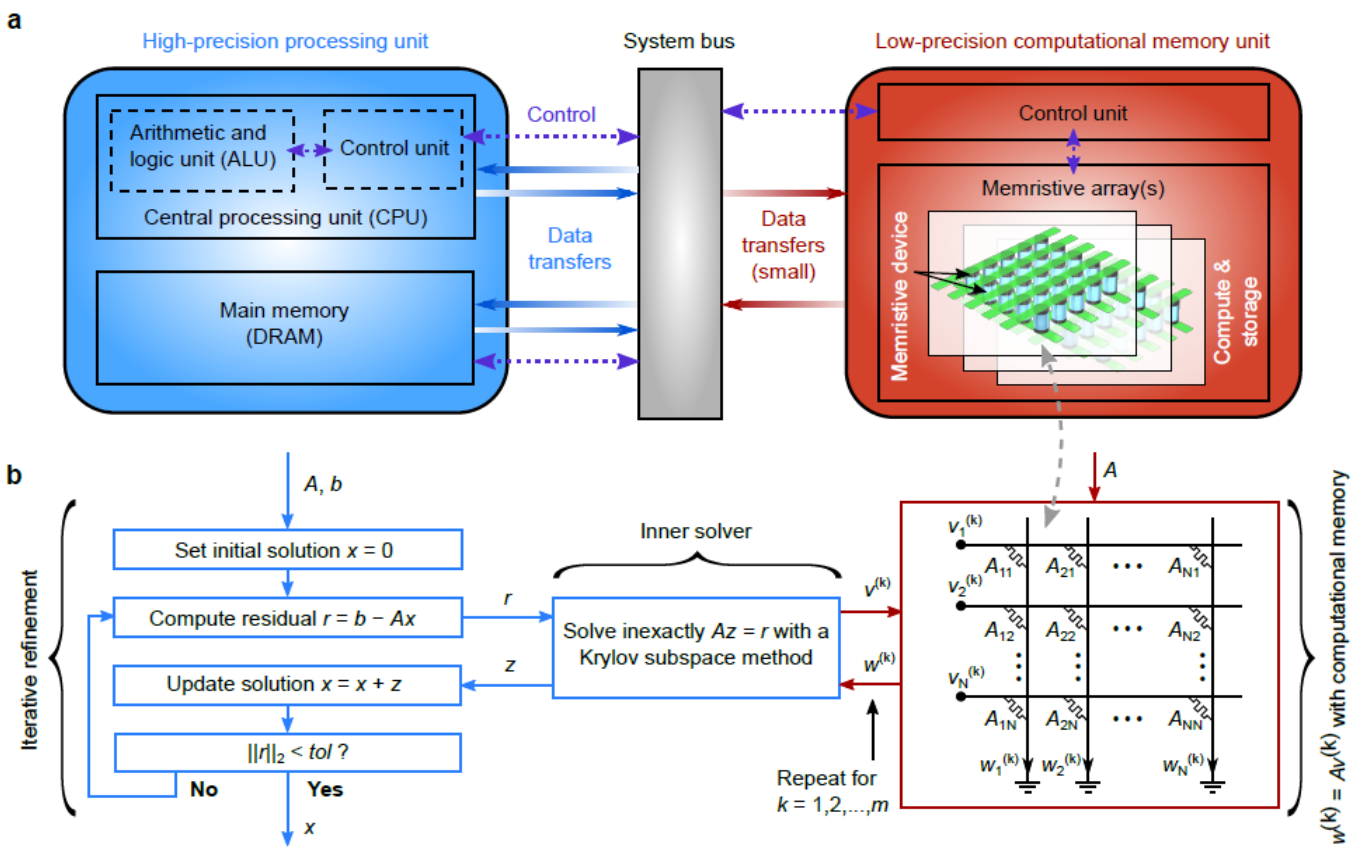

**Fig. S8: Mixed-precision in-memory computing implementation** of a linear equation solver according to ref. [18]. The problem is to find the solution x of Ax = b, where A is a NxN matrix and b is a N-dimensional vector.

In this application, matrix A is programmed once in the PCM array, and only read operations are performed subsequently in order to perform the matrix-vector multiplications to solve the linear system. Applications such as these require substantially smaller retention times than those needed for conventional memory/storage applications. To show this, we measured the runtime of the linear solver for different matrix sizes, using both an IBM POWER8 Central Processing Unit (CPU) and a NVIDIA P100 Graphical Processing Unit (GPU) as high-precision processing units. The linear solver was run for a tolerance of tol = $10^{-5}$. The PCM devices performing the matrix-vector multiplications were simulated with precision comparable to that achieved in the prototype chips used in the experiments presented in ref.[18]. As it can be seen in Supplementary Fig. S9, in the mixed-precision implementations for all matrix sizes the runtime is less than 1 second. Therefore, a minimum retention time for the PCM device of 1 s is sufficient for this particular application. Such retention time would be easily fulfilled by the 3 nm Sb devices investigated in the main manuscript.

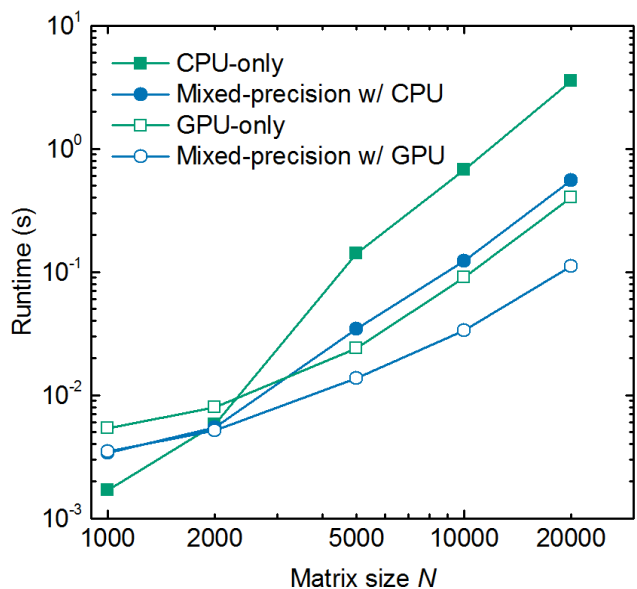

**Fig. S9: Measured runtime of linear solver for the mixed-precision implementation of Fig. S8 as well as CPU-only and GPU-only implementations.** In the mixed-precision implementations the runtime is less than 1 second for all matrix sizes investigated.

In the case of neuromorphic computing, certain computational tasks are performed using networks of neuronal and synaptic elements based on the computational principles of the brain. Phase-change memory (PCM) devices have been shown to capture both the neuronal and synaptic behavior in such an application [19,20]. The requirements for retention for such PCM devices are significantly different from those for conventional memory and storage applications. For example, in a PCM-based neuron, the PCM device is RESET every time the neuron fires. Hence, if we assume a rather conservative mean spiking frequency of 10 Hz, the average retention time required for the PCM device is 0.1 s. This is substantially lower than the retention times exhibited by the 3 nm Sb devices.

Even PCM-based synapses do not require a substantially long retention time. For example, an emerging application area is the use of PCM-based synapses for training deep neural networks [21]. In such an application, once the training is complete, the synaptic weights are copied and stored elsewhere. For each layer, the forward and backward propagation as well as the synaptic weight update can be performed in less than 1 $\mu$s [22]. For a network of 10 layers and a training set size of 100000, this translates to just 1 s of training time per epoch. Thus, a typical number of 20 training epochs would fit well within the retention time of the PCM devices presented in our work.